\documentclass[preprint, 12pt]{elsarticle}

\usepackage{amssymb}
\usepackage{amsmath}
\usepackage{lineno}
\usepackage{url}
\usepackage{bbm}
\usepackage[bb=libus]{mathalpha}

\journal{Physics letters B}

\begin{document}

\begin{frontmatter}



\title{From AdS Propagators to Celestial Propagators}


\author{Pongwit Srisangyingcharoen}
\ead{pongwits@nu.ac.th}
\affiliation{organization={The Institute for Fundamental Study, Naresuan University},
            city={Phisanulok},
            postcode={65000}, 
            country={Thailand}}

\begin{abstract}
In this paper, we investigate how AdS scalar propagators are represented in the celestial basis. Starting from the standard bulk-to-boundary propagator in Euclidean AdS space, we express the propagator in a Schwinger parametrization and construct the corresponding boundary-to-boundary propagator. We then transform the resulting propagators to the celestial basis using conformal primary wavefunctions for both massless and massive scalar fields. For the massless case, the celestial propagator reduces to an effectively two-dimensional boundary-to-boundary object on the celestial sphere dependent on the AdS/CFT conformal dimension $\Delta$ as a scaling factor. For the massive case, the celestial propagator exhibits a nontrivial kernel involving modified Bessel functions, closely resembling the momentum-space radial structure of AdS bulk-to-boundary propagators. The results suggest a structural translation from AdS propagators and celestial propagators.
\end{abstract}




\end{frontmatter}


\section{Introduction}\label{sec1}
The AdS/CFT correspondence has provided a profound connection between gravitational theories in anti-de Sitter (AdS) space and conformal field theories living on the boundary \cite{Maldacena:1997re,Witten:1998qj,Aharony:1999ti}. The correspondence has led to important insights across various areas of theoretical physics, including the thermodynamic properties of black holes \cite{Witten:1998zw,Strominger:1996sh,Ryu:2006bv,Hubeny:2007xt}, strongly coupled quantum field theories and quantum aspects of gravity \cite{Gubser:1998bc,Policastro:2001yc,Hartnoll:2009sz}, as well as string-theoretic descriptions of gauge dynamics and scattering amplitudes \cite{Polyakov:1998ju,Alday:2007hr, Penedones:2010ue,Fitzpatrick:2011ia, Srisangyingcharoen:2025ett}.

In this framework, bulk-to-boundary propagators play a central role since they encode how bulk fields are reconstructed from boundary data and determine the structure of correlation functions in the dual conformal field theory. Scalar propagators in AdS space have therefore been extensively studied both in position space and momentum space representations due to their importance in Witten diagrams and holographic computations \cite{Liu:1998ty,Costa:2014kfa, Gopakumar:2003ns, Gopakumar:2004qb}.

In parallel, celestial holography has emerged as a new approach toward flat space holography, where four-dimensional scattering amplitudes are recast as correlation functions on the two-dimensional celestial sphere. Rather than working in the conventional plane-wave basis, celestial amplitudes are constructed by transforming scattering amplitudes into a basis of conformal primary wavefunctions \cite{Pasterski:2017kqt,Pasterski:2017ylz}. This program is closely related to earlier studies of asymptotic symmetries, soft theorems, and infrared structures in gauge and gravitational theories \cite{Weinberg:1965nx, Strominger:2013jfa,He:2014laa, Kapec:2014opa}. In this formulation, Lorentz symmetry acts as the global conformal symmetry of the celestial sphere, leading to a two-dimensional conformal field theory interpretation of flat-space scattering processes. Considerable progress has been made in understanding various aspects of this holography such as celestial amplitudes, operator product expansions, conformal soft theorem, and celestial conformal field theories \cite{Arkani-Hamed:2020gyp, Stieberger:2018edy, Taylor:2023bzj, Donnay:2020guq, Puhm:2019zbl}.

An interesting aspect of celestial holography is that conformal primary wavefunctions themselves resemble bulk-to-boundary propagators. For massless particles, the conformal primary wavefunctions may be written as Mellin transforms of plane waves, while for massive particles they naturally involve bulk-to-boundary propagators on the three-dimensional hyperboloid $H_3$ \cite{Pasterski:2017kqt, Pasterski:2017ylz}. This structural similarity suggests possible deeper relations between AdS constructions and celestial amplitudes. 

Motivated by these observations, in this work we investigate how scalar propagators in AdS space are represented after transforming to the celestial basis. In particular, we focus on Euclidean AdS bulk-to-boundary propagators written in a Schwinger parametrization and examine their corresponding celestial representations for both massless and massive scalar fields. Starting from the standard AdS boundary-to-boundary propagator constructed by gluing two bulk-to-boundary propagators, we perform the celestial transform using conformal primary wavefunctions and study the resulting structures on the celestial sphere.

This paper is organized as follows. In Section 2, we review scalar bulk-to-boundary propagators in Euclidean AdS space and rewrite them using Schwinger parametrization. In Section 3, we review conformal primary wavefunctions and celestial amplitudes for both massless and massive scalar fields. In Section 4, we derive the celestial representation of the massless AdS boundary-to-boundary propagator and identify its effective two-dimensional structure on the celestial sphere. In Section 5, we extend the analysis to the massive case and obtain the corresponding celestial propagator. Finally, Section 6 contains conclusions.

\section{AdS Scalar propagator}\label{sec2}
In this section, we would like to exhibit the bulk-to-boundary propagators in $\text{AdS}_{d+1}$ in a Schwinger representation for a massive scalar field. It is convenient to work in the Poincar\'e coordinates of Euclidean $\text{AdS}_{d+1}$,
\begin{equation}
    ds^2=L^2\frac{d\mathbf{z}^2+dz_0^2}{z_0^2}
\end{equation}
where $L$ is called the AdS curvature radius. Let us consider the massive scalar action
\begin{equation}
    S[\phi]=\frac{1}{2}\int d^d\mathbf{z} dz_0 \sqrt{g}\left(g^{MN}\partial_M\phi(z)\partial_N\phi(z)+m^2\phi^2(z) \right)
\end{equation} 
where $\phi(z)=\phi(z_0,\mathbf{z})$. A bulk-to-boundary propagator $K(z_0,\mathbf{z};\mathbf{z}')$ can be obtained from the wave equation
\begin{equation}
    \left(z_0^{d+1}\frac{\partial}{\partial z_0}\left(z_0^{-d+1}\frac{\partial}{\partial z_0}  \right)+z_0^2\Box-m^2L^2\right)K=0. \label{bulktoboundary prop}
\end{equation}
where $\Box$ is $d$-dimensional Laplacian operator in direction $\mathbf{z}$. It is well-known that the expression for bulk-to-boundary propagator takes the form 
\begin{equation}
    K_{\Delta}(z_0,\mathbf{z};\mathbf{z}')=\frac{\Gamma(\Delta)}{\pi^{d/2}\Gamma(\Delta-\frac{d}{2})}\left( \frac{z_0}{z_0^2+(\mathbf{z}-\mathbf{z}')^2}\right)^\Delta
\end{equation}
where $\Delta=\frac{1}{2}\left(d+\sqrt{d^2+4m^2L^2}\right)$ \cite{Witten:1998qj}. The AdS bulk field $\phi(z)$ is related to the boundary data by
\begin{equation}
    \phi(z_0,\mathbf{z})=\int d^d\vec{x} \ K_\Delta(z_0,\mathbf{z};\mathbf{x})\phi_0(\mathbf{x}).
\end{equation}

Using a Schwinger parametrization, one obtains
\begin{equation}
    K(z_0,\mathbf{z};\mathbf{z}')=\frac{1}{z_0^\Delta\pi^{d/2}\Gamma(\Delta-\frac{d}{2})}\int_0^\infty d\rho \ \rho^{\Delta-1}e^{-\rho}e^{-\frac{(\mathbf{z}-\mathbf{z}')^2}{z_0^2}\rho}.
\end{equation}
We can then construct a boundary-to-boundary propagator through
\begin{align}
   K(\mathbf{x},\mathbf{y})=&\frac{1}{2}\int_0^\infty \frac{dt}{t^{\frac{d}{2}+1}}\int d^dz \ K_\Delta(t,\mathbf{z};\mathbf{y})K_\Delta(t,\mathbf{z};\mathbf{x}) \label{boundary to boundary prob}  
\end{align}
which can be expressed as
\begin{align}
   \frac{1}{2\pi^d}&\frac{1}{(\Gamma(\Delta-d/2))^2}\int_0^\infty \frac{dt}{t^{\Delta+\frac{d}{2}+1}}\int d^dz \int_0^\infty d\rho_1 \int_0^\infty d\rho_2 \nonumber \\
    &\rho_1^{\Delta-1}\rho_2^{\Delta-1} e^{-(\rho_1+\rho_2)} e^{-(\mathbf{z}-\mathbf{x})^2\frac{\rho_1}{t}}e^{-(\mathbf{z}-\mathbf{y})^2\frac{\rho_2}{t}}
\end{align}
where $t=z_0^2$.

Alternatively, one may utilize the heat kernel,
\begin{equation}
    \langle \mathbf{x}|e^{\tau\Box}|\mathbf{y}\rangle=\frac{1}{(4\pi \tau)^{d/2}}e^{-(\mathbf{x}-\mathbf{y})^2/4\tau}
\end{equation}
to rewrite
\begin{equation}
    K_\Delta(t,\mathbf{z};\mathbf{z}')=\frac{t^{(d-\Delta)/2}}{\Gamma(\Delta-\frac{d}{2})}\int_0^\infty d\rho \ \rho^{\Delta-\frac{d}{2}-1} e^{-\rho} \langle \mathbf{z}|e^{\frac{t}{4\rho}\Box}|\mathbf{z}'\rangle.
\end{equation}
This gives 
\begin{align}
   K(\mathbf{x},\mathbf{y})=&\frac{1}{2(\Gamma(\Delta-d/2))^2}\int_0^\infty \frac{dt}{t^{\Delta+1-\frac{d}{2}} } \int d^dz\int_0^\infty d\rho_1 \int_0^\infty d\rho_2 \nonumber \\
   &\times \ (\rho_1\rho_2)^{\Delta-\frac{d}{2}-1} e^{-(\rho_1+\rho_2)}\langle \mathbf{y}|e^{\frac{t}{4\rho_2}\Box}|\mathbf{z}\rangle\langle \mathbf{z}|e^{\frac{t}{4\rho_1}\Box}|\mathbf{x}\rangle. \label{2-pt ads prop}
\end{align}
The expression describes the standard boundary-to-boundary propagator constructed from two identical bulk-to-boundary propagators of weight $\Delta$. In this paper, we examine how such an AdS propagator is perceived in the 2D celestial space. Therefore, throughout the paper, we work in $d=4$ dimensions. 

\section{Conformal primary wavefunctions and celestial amplitudes}

In a two-dimensional celestial CFT, celestial amplitudes are scattering amplitudes expanded on the basis of conformal primary wavefunctions \cite{Pasterski:2016qvg, Pasterski:2017kqt} rather than in the usual plane-wave basis. A conformal primary wavefunction $\phi_{h_i}^{\pm}(\mathbb{z}_i,\mathbf{x}_i)$ with the conformal dimension $h_i$ can be thought of as a bulk-to-boundary propagator in $\mathbb{R}^{3,1}$ that maps four-dimensional Minkowski spacetime $\mathbf{x}_i$ to a boundary point $\mathbb{z}_i=(z_i,\bar{z}_i)$ on the celestial sphere. 

Consider a $4D$ scattering amplitude of $n$-incoming and $m-$outgoing particles, $A_{n+m}(\mathbf{x}_i)$. The corresponding celestial amplitude $\mathcal{A}_{n+m}(h_i,z_i)$ is defined through
\begin{align}
    \mathcal{A}_{n+m}(h_i,\mathbb{z}_i)=&\int \prod_{i=1}^{n+m} d^4x_i \left( \prod_{j=1}^n \phi^+_{h_j}(\mathbb{z}_j,\mathbf{x}_j) \right) \nonumber \\
    &\times\left( \prod_{k=1}^m \phi^-_{h_j}(\mathbb{z}_k,\mathbf{x}_k) \right) \, A_{n+m}(\mathbf{x}_i). \label{celestial trans}
\end{align}
The plus and minus signs of the conformal primary wavefunctions denote incoming and outgoing scattering waves, respectively. Some may write the conformal dimensions $h_j=1+i\lambda_j$, $\lambda_j\in\mathbb{R}$.

The four-momentum $k^\mu$ of a scattering particle is parametrized according to its mass. For a massless particle, 
\begin{align}
    \mathbf{k}^{\mu} &= \omega \mathbf{q}^{\mu}(\mathbb{z})\nonumber \\
    &= \omega_j \bigl(
    1+|z|^2,\,
    z+\bar{z},\,
    -i(z-\bar{z}),\,
    1-|z|^2
    \bigr), \label{4 momentum massless}
\end{align}
where $\omega_j$ is the angular frequency associated with the energy of the external particle and $\mathbf{q}^{\mu}_j$ is a  null vector pointing toward a point $\mathbb{z}=(z,\bar{z})$ on the celestial sphere. For a massive particle with mass $m$, the momentum is 
\begin{align}
    \mathbf{k}^{\mu} &= m \mathbf{p}^{\mu}(y,\mathbb{z}) \nonumber \\
    &=\frac{m}{2y}\bigl(1+y^2+|z|^2,z+\bar{z},-i(z-\bar{z}),1-y^2-|z|^2\bigr) \label{4 momentum massive}
\end{align}
where $\mathbf{p}^\mu$ represents a unit vector directed toward a point on the three-dimensional upper unit hyperboloid $H_3$ with $y$ denoting the radial distance into the bulk of $H_3$. The complex coordinates $\mathbb{z}=(z,\bar{z})$ identify the location on the celestial sphere where the particle's trajectory intersects the boundary.

The primary conformal wavefunctions for massless and massive scalars are
\begin{align}
    \phi^\pm_{h}(\mathbb{z},\mathbf{x}^\mu)=&\int_0^\infty d\omega \, \omega^{h-1} e^{\pm i\omega \mathbf{q}(\mathbb{z})\cdot \mathbf{x}} \\
    \phi^\pm_{h}(\mathbb{z},\mathbf{x}^\mu)=&\int_0^\infty \frac{dy}{y^3} \int d^2w G_h(y,\mathbb{w};\mathbb{z}) e^{\pm im\mathbf{p}(y,\mathbb{w})\cdot\mathbf{x}}
\end{align}
respectively. $G_h(y,\mathbb{w};\mathbb{z})$ is the $4D$-to-$2D$ bulk-to-boundary propagator in Minkowski spacetime whose expression is 
\begin{equation}
    G_h(\mathbf{{p}}(y,\mathbb{w});\mathbf{q}(\mathbb{z}))=\frac{1}{(-\mathbf{p}\cdot\mathbf{q})^h}=\left( \frac{y}{y^2+\vert w-z \vert^2}\right)^h. \label{minkowski bulk to boundary prop}
\end{equation}


The conformal primary wavefunctions also admit a shadow
transformation analogous to that in ordinary conformal field
theory \cite{Ferrara:1972uq}. In a two-dimensional conformal field theory,
a primary operator with conformal dimension $h$ is related to
its shadow operator with conformal dimension $2-h$. For the
celestial conformal primary wavefunctions, the shadow trans
formation acts as an integral transform on the celestial sphere,
\begin{equation}
    \widetilde{\phi}^{\pm}_{h}(\mathbb{z},\mathbf{x})
    =\frac{\Gamma(h)}{\pi\Gamma(h-1)}
    \int d^2w
    \frac{
        \phi^{\pm}_{h}(\mathbb{w},\mathbf{x})
    }{
        |z-w|^{2(2-h)}
    } .\label{shadow transform}
\end{equation}
For scalar conformal primary wavefunctions, the shadow
transform produces another conformal primary wavefunction
with shadow conformal dimension
\begin{equation}
    h \rightarrow 2-h.
\end{equation}
Accordingly, the shadow transform of the conformal
primary wavefunction (15) is 
\begin{equation}
    \widetilde{\phi}^{\pm}_{h}(\mathbb{z},\mathbf{x})
   =
    \phi^{\pm}_{2-h}(\mathbb{z},\mathbf{x})
\end{equation}
for both scalar massless and massive cases.

Beyond the standard Mellin transform over the energy variables,
alternative integral representations of celestial amplitudes have
also been explored. Recently, a Mellin-like transformation acting
directly on the celestial coordinates $(z_i,\bar z_i)$ was introduced for celestial massless amplitudes \cite{Yuenyong:2026wnv}. Inspired
by closed-string worldsheet integrals, the transformation maps
$(z_i,\bar z_i)$ to a new set of complex variables
$(s_i,\bar s_i)$ and leads to constraints associated with global
conformal invariance.

\section{From Massless boundary-to-boundary propagator to celestial propagator}
Since the boundary-to-boundary propagator $K(\mathbf{x},\mathbf{y})$ is in a flat four-dimensional space, we can map this to the celestial sphere. Let first consider a massless scalar field with AdS weight $\Delta$. According to (\ref{celestial trans}), the celestial propagator for a massless particle reads
\begin{align}
     \mathcal{K}(h_i,z_i,\bar{z}_i)= &\int_0^\infty d\omega_1\int_0^\infty d\omega_2 \, \omega_1^{h_1-1}\omega_2^{h_2-1} \nonumber \\
     &\times\int d^4x d^4y \ K(\mathbf{x},\mathbf{y}) \ e^{i\omega_1\mathbf{q}_1(\mathbb{z}_1)\cdot \mathbf{x}-i\omega_2\mathbf{q}_2(\mathbb{z}_2)\cdot \mathbf{y}}. \label{celestial prop1}
\end{align}
The second line of the above equation is simply the Fourier transform of the propagator $K(\mathbf{x},\mathbf{y})$, by which we will refer to it as $K(\omega_i,z_i,\bar{z}_i)$. Performing Gaussian integration, one obtains
\begin{align}
    K(\omega_i,z_i,\bar{z}_i)=&\frac{(2\pi)^4}{2}\delta^{(4)}(\omega_1\mathbf{q}_1-\omega_2\mathbf{q}_2)\int_0^\infty\frac{dt}{t^{\Delta-1}} \int_0^\infty d\rho_1 \int_0^\infty d\rho_2 \nonumber \\
    &\times\rho_1^{\Delta-3}\rho_2^{\Delta-3} e^{-(\rho_1+\rho_2)} e^{-\omega_1^2q_1^2\frac{t}{4\rho_1}}e^{-\omega_2^2q_2^2\frac{t}{4\rho_2}}.\label{celestial prop2}
\end{align}
The conservation of momentum together with their massless representations  (\ref{4 momentum massless}) allows us to write
\begin{align}
\omega_1^2q_1^2=\omega_2^2q_2^2=\omega_1\omega_2\mathbf{q}_1\cdot \mathbf{q}_2=2\omega_1\omega_2|z_{12}|^2
\end{align}
where $z_{12}=z_1-z_2$. Note that we work in the mostly
minus Minkowski signature. Accordingly, the expression (\ref{celestial prop2}) becomes
\begin{align}
\frac{(2\pi)^4}{2}&\delta^{(4)}(\omega_1\mathbf{q}_1-\omega_2\mathbf{q}_2)\int_0^\infty\frac{dt}{t^{\Delta-1}} \int_0^\infty d\rho_1 \int_0^\infty d\rho_2 \nonumber \\
    &\times\rho_1^{\Delta-3}\rho_2^{\Delta-3} e^{-(\rho_1+\rho_2)} e^{-\frac{\omega_1\omega_2}{2}|z_{12}|^2t\left(\frac{1}{\rho_1}+\frac{1}{\rho_2} \right)}. \label{massless celes prop}
\end{align}

The Mellin transformation then converts the momentum space propagator to the corresponding two-dimensional celestial one through
\begin{equation}
    \mathcal{K}(h_i,z_i,\bar{z}_i)=\int_0^\infty d\omega_1\int_0^\infty d\omega_2 \, \omega_1^{h_2-1}\omega_2^{h_2-1} K(\omega_i,z_i,\bar{z}_i).
\end{equation}
Integration over $\omega_i$ is not straightforward as the integrating variables appear inside the Dirac delta function. To this extent, we decompose the delta function as 
\begin{equation}
    \delta^{(4)}(\mathbf{k} - \mathbf{k}')
    = \frac{1}{\omega}\,
    \delta(\omega - \omega')\,
    \delta^{(2)}(z - z')\,
    \delta(k^2)\, \Theta(k^0). \label{delta decom}
\end{equation}
More detail is shown in the appendix \ref{dirac del}. We may ignore the step function as $k^0$ is always non-negative.

First let us use the fact that $z_1=z_2$ and $\bar{z}_1=\bar{z}_2$ via the constraint $\delta^{(2)}(z_{12})$. This greatly simplifies the celestial propagator as 
\begin{equation}
\exp{\left(-\frac{\omega_1\omega_2}{2}|z_{12}|^2t\left(\frac{1}{\rho_1}+\frac{1}{\rho_2}\right)\right)}=1.     
\end{equation}
Therefore,
\begin{align}
    \mathcal{K}(h_i,z_i,\bar{z}_i)=&\frac{(2\pi)^4}{2}\delta(k^2_1)\int_0^\infty d\omega_1\int_0^\infty d\omega_2 \, \omega_1^{h_1-2}\omega_2^{h_2-1} \delta(\omega_1-\omega_2) \nonumber \\
    &\times \delta^{(2)}(z_1-z_2)\int_0^\infty\frac{dt}{t^{\Delta-1}} \int_0^\infty d\rho_1 \int_0^\infty d\rho_2 (\rho_1\rho_2)^{\Delta-3} e^{-(\rho_1+\rho_2)} \nonumber \\
    =&\frac{(2\pi)^4}{2}\delta(k^2)(\Gamma(\Delta-2))^2\delta^{(2)}(z_1-z_2)\int_0^\infty\frac{dt}{t^{\Delta-1}} \int_0^\infty d\omega \, \omega^{h_1+h_2-3}.
\end{align}
The factor $\delta(k^2)$ is  divergent on-shell providing a divergent prefactor which is considered as an IR divergence. The $\omega$-integral provides a constraint
\begin{equation}
    \int_0^\infty d\omega \, \omega^{h_1+h_2-3}=2\pi \delta(\lambda_1+\lambda_2)
\end{equation}
where $h_j=1+1\lambda_j$. This recovers a known condition for celestial amplitudes. Accordingly, up to an overall divergent normalization factor, the celestial propagator becomes
\begin{equation}
    \mathcal{K}(h_i,z_i,\bar{z}_i) = \mathcal{N}(\Gamma(\Delta-2))^2 \int_0^\infty\frac{dt}{t^{\Delta-1}}  \delta(\lambda_1+\lambda_2)\delta^{(2)}(z_1-z_2).\label{celes prob}
\end{equation}
Although we could integrate out the AdS bulk radius coordinate $t$, leaving it untouched allows us  to relate the celestial two-point function with a bulk-to-boundary propagator in AdS space. 

To match with the propagator (\ref{boundary to boundary prob}), we can decompose the delta function to be
\begin{equation}
    \delta^{(2)}(z_1-z_2)=\int d^2z \delta^{(2)}(z_1-z)\delta^{(2)}(z-z_2).
\end{equation}
Accordingly, the celestial propagator takes the form
\begin{align}
\mathcal{K}(\Delta,z_i,\bar z_i)
=
\mathcal{N}
\int_0^\infty \frac{dt}{t^3}
\int d^2z\,
\mathcal{K}_{\Delta}(t;\mathbb{z}_1,\mathbb{z})
\mathcal{K}_{\Delta}(t;\mathbb{z}_2,\mathbb{z}),
\label{celestial prop3}
\end{align}
where
\begin{equation}
\mathcal{K}_{\Delta}(t;\mathbb{z}_i,\mathbb{z})
=
\frac{\Gamma(\Delta-2)}
{t^{\frac{\Delta}{2}-2}}
\delta^{(2)}(z_i-z).
\label{massless celes prop final}
\end{equation}
We may interpret
$\mathcal{K}_{\Delta}(t;z_i,z)$ as a scaling celestial propagator
that relates the celestial point $z_i$ to the intermediate point $z$
on the celestial sphere. The propagator is characterized by a scaling
dependence on the AdS radial parameter $t$. However, the presence of the
Dirac delta function makes this kernel distributional and prevents it
from taking the conventional non-local form of a conformal two-point
function. Following the rectified celestial dictionary of
Furugori et al.~\cite{Furugori:2023hgv}, this distributional form can be
converted into a conformally covariant power-law kernel by incorporating
the shadow transformation.

We therefore apply the shadow transformation to the celestial endpoints
$z_1$ and $z_2$. For a scalar celestial operator of conformal dimension
$h_i$, the shadow transformation gives
\begin{align}
\widetilde{\mathcal{K}}_{\Delta,h_i}(t;z_i,z)
&=
\frac{h_i-1}{\pi}
\int d^2w_i\,
\frac{
\mathcal{K}_{\Delta}(t;\mathbb{w}_i,\mathbb{z})
}{
|z_i-w_i|^{2(2-h_i)}
}
\nonumber\\
&=
\frac{h_i-1}{\pi}
\frac{\Gamma(\Delta-2)}
{t^{\frac{\Delta}{2}-2}}
\frac{1}{
|z_i-z|^{2(2-h_i)}
}.
\label{shadow-celestial-kernel}
\end{align}
The shadow transformation thus maps the distributional localization at
$z_i=z$ into a non-local power-law dependence on the celestial
separation $|z_i-z|$.

The corresponding rectified celestial propagator is obtained by
substituting the shadow-transformed kernels into the gluing representation
in ~\eqref{celestial prop3}. Namely,
\begin{align}
\widetilde{\mathcal{K}}
(\Delta,h_i,z_i,\bar z_i)
=
\mathcal{N}
\int_0^\infty \frac{dt}{t^3}
\int d^2z\,
\widetilde{\mathcal{K}}_{\Delta,h_1}(t;\mathbb{z}_1,\mathbb{z})
\widetilde{\mathcal{K}}_{\Delta,h_2}(t;\mathbb{z}_2,\mathbb{z}).
\label{rectified celestial prop}
\end{align}
It is important to distinguish the two representations. The original
expression in ~\eqref{celestial prop3} is the distributional form
obtained directly from the Mellin transform, whereas ~\eqref{rectified celestial prop} is its shadow-transformed
representation. The latter provides a conformally covariant realization
of the two-point correlation function.

The structure of ~\eqref{rectified celestial prop} can now be compared
directly with the AdS boundary-to-boundary propagator in
~\eqref{boundary to boundary prob}. In the latter, the boundary-to-
boundary propagator is constructed by gluing two bulk-to-boundary
propagators through an integration over the common bulk point. The
rectified celestial propagator possesses precisely the same gluing
structure: the two celestial kernels are attached to a common
intermediate point $z$ and integrated over the radial parameter $t$ and
the intermediate celestial coordinate.

This structural similarity suggests the correspondence
\begin{equation}
K_{\Delta}(t,\mathbf{z};\mathbf{x}_i)
\quad\longrightarrow\quad
\widetilde{\mathcal{K}}_{\Delta,h_i}(t;\mathbb{z},\mathbb{z}_i).
\label{bulk boundary celestial translation}
\end{equation}
Thus, this provides a celestial
analogue of the AdS bulk-to-boundary propagator. In this correspondence,
the AdS radial scaling is retained through the factor
$t^{-\frac{\Delta}{2}+2}$, inherited from the original bulk-to-boundary
propagator, while the dependence on the external celestial point is
encoded in the conformally covariant factor
$|z-z_i|^{-2(2-h_i)}$.

It is important to emphasize that the correspondence in
~\eqref{bulk boundary celestial translation} is structural rather
than an identification of the underlying coordinates. The celestial
coordinates $\mathbb{z}$ are not the four-dimensional boundary
coordinates $\mathbf{x}^{\mu}$. Instead, they parametrize the direction
of null momenta in the momentum-space representation underlying the
celestial Mellin transform.

\section{From Massive boundary-to-boundary propagator to celestial propagator}
Similar to the previous section, we begin with the boundary-to-boundary propagator $K(\mathbf{x},\mathbf{y})$ and map it to the corresponding celestial propagator using (\ref{celestial trans}),
\begin{align}
    \mathcal{K}(h_i,z_i,\bar{z}_i)=& \int_0^\infty \frac{dy_1}{y_1^3}\int d^2w_1 \int_0^\infty\frac{dy_2}{y_2^3}\int d^2w_2 \nonumber \\ &\times G_{h_1}(y_1,\mathbb{w}_1;\mathbb{z}_1)G_{h_2}(y_2,\mathbb{w}_2;\mathbb{z}_2) K(y_i,w_i,\bar{w}_i) \label{massive celestial prop}
\end{align}
where
\begin{equation}
    K(y_i,w_i,\bar{w}_i)=\int d^4 xd^4y \ K(\mathbf{x},\mathbf{y}) \ e^{im\mathbf{p}_1(y_1,\mathbb{w}_1)\cdot \mathbf{x}-im\mathbf{p}_2(y_2,\mathbb{w}_2)\cdot \mathbf{y}}.
\end{equation}
First, we examine the propagator in momentum space $K(y_i,w_i,\bar{w}_i)$. By performing the Gaussian integral, we obtain
\begin{align}
K(y_i,w_i,\bar{w}_i)=&\frac{(2\pi)^4}{2}\delta^{4}(m\mathbf{p}_1-m\mathbf{p}_2)\int_0^\infty\frac{dt}{t^{\Delta-1}} \int_0^\infty d\rho_1 \int_0^\infty d\rho_2 \nonumber \\
    &\times(\rho_1\rho_2)^{\Delta-3} e^{-(\rho_1+\rho_2)} e^{-m^2p_1^2\frac{t}{4\rho_1}}e^{-m^2p_2^2\frac{t}{4\rho_2}}.\label{fourier prop}
\end{align}

Naively, if we apply the on-shell relations $p_1^2=p_2^2=1$, the whole integration (\ref{fourier prop}) would be independent of $(y_i,w_i,\bar{w}_i)$. Therefore, we may write
\begin{equation}
    K(z_i,\bar{z}_i)=f(\Delta)\delta^{4}(\mathbf{p}_1-\mathbf{p}_2)
\end{equation}
for some function $f(\Delta)$ that depends only on $\Delta$.
 Substituting into (\ref{massive celestial prop}), it yields
\begin{align}
    \mathcal{K}(h_i,z_i,\bar{z}_i)=& f(\Delta)\int_0^\infty \frac{dy_1}{y_1^3}\int d^2w_1 \int_0^\infty\frac{dy_2}{y_2^3}\int d^2w_2 \nonumber \\
    &\times 4y_1^3\delta(y_1 - y_2)\,
    \delta^{(2)}(w_{12}) \delta(p^2-1)\Theta(p^0) \nonumber \\ &\times G_{h_1}(y_1,\vec{w}_1;\vec{z}_1)G_{h_2}(y_2,\vec{w}_2;\vec{z}_2) \label{massive celestial prop2}
\end{align}
where the Dirac delta function $\delta^{4}(\mathbf{p}_1-\mathbf{p}_2)$ was decomposed. More details on decomposition can be found in  \ref{dirac del}. Upon reparameterizing the bulk-to-boundary propagator $G_{h_i}(y_i,\mathbb{w}_i;\mathbb{z}_i)$ as
\begin{equation}
    \frac{1}{y_i^{h_i}\Gamma(h_i)}\int_0^\infty d\alpha_i \, \alpha^{h_i-1}e^{-\alpha_i} e^{-\frac{|w_i-z_i|^2}{y_i^2}\alpha_i}
\end{equation}
and integrating over the variables $y_2$ and $w_2$, one finds
\begin{align}
    \mathcal{K}(h_i,z_i,\bar{z}_i)=& C'f(\Delta)\int_0^\infty \frac{dy}{y^3} y^{-(h_1+h_2)}\iint_0^\infty d\alpha_1 d\alpha_2 \, e^{-(\alpha_1+\alpha_2)}\nonumber \\
    &\times\alpha_1^{h_1-1}\alpha_2^{h_2-1}\left(\int d^2w \,e^{-\frac{1}{y^2}\left( |w-z_1|^2\alpha_1+|w-z_2|^2\alpha_2 \right)}\right). \label{massive celestial prop3}
\end{align}
Take into account the Gaussian integral inside the parentheses,
by which we can rewrite its exponent $\alpha_1|w-z_1|^2+\alpha_2|w-z_2|^2$ as
\begin{equation}
(\alpha_1+\alpha_2)\left\vert w-\frac{\alpha_1z_1+\alpha_2z_2}{\alpha_1+\alpha_2} \right\vert^2+\frac{\alpha_1\alpha_2}{\alpha_1+\alpha_2}\left\vert z_{12} \right\vert^2.
\end{equation}
Performing the two-dimensional Gaussian integral, it yields
\begin{align}
    \int d^2w \,e^{-\frac{1}{y^2}\left( |w-z_1|^2\alpha_1+|w-z_2|^2\alpha_2 \right)}=\frac{\pi y^2}{\alpha_1+\alpha_2}e^{-\frac{\alpha_1\alpha_2}{\alpha_1+\alpha_2}\frac{|z_{12}|^2}{y^2}}.
\end{align}
Put the above result into (\ref{massive celestial prop3}). It yields
\begin{align}
    \mathcal{K}(h_i,z_i,\bar{z}_i)=& \pi C'f(\Delta)\iint_0^\infty d\alpha_1 d\alpha_2 \frac{\alpha_1^{h_1-1}\alpha_2^{h_2-1}}{\alpha_1+\alpha_2} e^{-(\alpha_1+\alpha_2)}\nonumber \\
    &\int_0^\infty dy \, y^{-(h_1+h_2)-1}e^{-\frac{\alpha_1\alpha_2}{\alpha_1+\alpha_2}\frac{|z_{12}|^2}{y^2}}. 
\end{align}
One can turn the $y$-integral in the second line into the Gaussian integral by changing $y \rightarrow \frac{1}{y}$. Accordingly, the above equation takes the form
\begin{align}
    \mathcal{K}(h_i,z_i,\bar{z}_i)=& \frac{\pi}{2} C'f(\Delta)\Gamma\left(\frac{h_1+h_2}{2} \right)\frac{1}{|z_{12}|^{h_1+h_2}}\nonumber \\
    &\iint_0^\infty d\alpha_1 d\alpha_2 \frac{\alpha_1^{\frac{h_1-h_2}{2}-1}\alpha_2^{-\frac{h_1-h_2}{2}-1}}{(\alpha_1+\alpha_2)^{1-\frac{h_1+h_2}{2}}} e^{-(\alpha_1+\alpha_2)}.
\end{align}
What remains is to evaluate the $\alpha-$integral in the second line.

To continue, we change the variables $\alpha_1=xu$ and $\alpha_2=x(1-u)$ with $x\in (0,\infty)$ and $u\in(0,1)$. Hence, we can write the integration with respect to the variable $u$ as
\begin{align}
    \int_0^1 du \, u^{\frac{h_1-h_2}{2}-1}(1-u)^{-\frac{h_1-h_2}{2}-1}
\end{align}
which is equal to 
\begin{equation}
    2\int_{-\infty}^\infty dv \, e^{v(h_1-h_2)}=4\pi\delta(\lambda_1-\lambda_2).
\end{equation}
This is the conservation of conformal dimensions for massive two point correlations.


Despite the desirable result, the expression still lacks a structure resembling bulk-to-boundary propagators. To this extent, let us reconsider the form of the propagator (\ref{fourier prop}) but instead of using $p_i^2=1$, we use the conservation of momentum to rewrite
\begin{equation}
    p_1^2=p_2^2=\mathbf{p}_1\cdot \mathbf{p}_2=\frac{|w_{12}|^2+y_1^2+y_2^2}{2y_1y_2}
\end{equation}
where $w_{12}=w_1-w_2$. Consequently, the celestial propagator (\ref{massive celestial prop}) becomes
\begin{align}
    \mathcal{K}(h_i,z_i,\bar{z}_i)=&N \int_0^\infty \frac{dy_1}{y_1^3}\int d^2w_1 \int_0^\infty\frac{dy_2}{y_2^3}\int d^2w_2 \nonumber \\ 
    & y_1^3\delta(y_1 - y_2)\,
    \delta^{(2)}(w_{12})G_{h_1}(y_1,\mathbb{w}_1;\mathbb{z}_1) \nonumber  \\
    &\times G_{h_2}(y_2,\mathbb{w}_2;\mathbb{z}_2)\int_0^\infty\frac{dt}{t^{\Delta-1}} \iint_0^\infty d\rho_1  d\rho_2  \nonumber\\
    &\times (\rho_1\rho_2)^{\Delta-3}e^{-(\rho_1+\rho_2)}e^{-m^2\frac{|w_{12}|^2+y_1^2+y_2^2}{2y_1y_2}t\left(\frac{1}{\rho_1}+\frac{1}{\rho_2}\right)} \label{massive celes porpp}
\end{align}
with some constant $N$. Integrating over $y_2$ and $w_2$, one obtains
\begin{align}
\mathcal{K}(h_i,z_i,\bar{z}_i)=&N\int_0^\infty \frac{dy}{y^3}\int d^2w \int_0^\infty\frac{dt}{t^{\Delta-1}}\nonumber \\ 
    &\times \left( \frac{y}{y^2+|w-z_1|^2}\right)^{h_1} \left( \frac{y}{y^2+|w-z_2|^2}\right)^{h_2} \nonumber  \\
    &\times \iint_0^\infty d\rho_1  d\rho_2 \, (\rho_1\rho_2)^{\Delta-3}e^{-(\rho_1+\rho_2)}e^{-m^2t\left(\frac{1}{\rho_1}+\frac{1}{\rho_2}\right)}. \label{massive celestial prop4}
\end{align}
The $\rho-$integral matches the standard integral representation of the modified Bessel function of the second kind \cite{Gradshteyn:1943cpj} (See eq.3.471.9 thereof) which is
\begin{equation}
    \int_0^\infty x^{\nu-1}e^{-\frac{\beta}{x}-\gamma x}dx=2\left( \frac{\beta}{\gamma}\right)^{\nu/2}K_\nu(2\sqrt{\beta\gamma}). \label{Bessel}
\end{equation}
Consequently, the massive celestial propagator takes the form
\begin{align}
\mathcal{K}(h_i,z_i,\bar{z}_i)=&N\int_0^\infty \frac{dy}{y^3}\int d^2w \int_0^\infty\frac{dt}{t} \, \left( 2m^{(\Delta-2)}K_{\Delta-2}(2m\sqrt{t})\right)^2\nonumber \\ 
    &\times \left( \frac{y}{y^2+|w-z_1|^2}\right)^{h_1} \left( \frac{y}{y^2+|w-z_2|^2}\right)^{h_2}.
\label{massive celestial prop5}
\end{align}
Therefore, we may introduce a celestial bulk-to-boundary propagator 
\begin{equation}
    \widetilde{\mathcal{K}}_{\Delta,h}^m(t;y,\mathbb{w};\mathbb{z})=\left( \frac{y}{y^2+|w-z|^2}\right)^{h}\frac{1}{t^{\frac{\Delta}{2}-2}}\left( 2(m\sqrt{t})^{\Delta-2}K_{\Delta-2}(2m\sqrt{t})\right) \label{massive celes prop}
\end{equation}
such that
\begin{align}
    \mathcal{K}(h_i,z_i,\bar{z}_i)=&N\int_0^\infty \frac{dy}{y^3}\int_0^\infty\frac{dt}{t^3}\int d^2w \nonumber \\
    &\times \widetilde{\mathcal{K}}_{\Delta,h_1}^m(t;y,\mathbb{w};\mathbb{z}_1)\widetilde{\mathcal{K}}_{\Delta,h_2}^m(t;y,\mathbb{w};\mathbb{z}_2). \label{massive celestial prop-final}
\end{align}
Note that when the mass $m$ approaches zero, the factor
\begin{equation}
     2(m\sqrt{t})^{\Delta-2}K_{\Delta-2}(2m\sqrt{t}) \rightarrow \Gamma(\Delta-2)
\end{equation}
giving the same  factor as in the massless propagator (\ref{massless celes prop final}).

An appearance of the modified Bessel function of the second kind $K_{\Delta-2}(2m\sqrt{t})$ inside the celestial bulk-to-boundary propagator suggests a remnant of the momentum-space AdS bulk-to-boundary propagator after transforming to the celestial basis. 

Comparing (\ref{massive celestial prop-final}) with the AdS boundary-to-boundary propagator in
(\ref{boundary to boundary prob}), we find that the two expressions possess similar
structure. In (\ref{boundary to boundary prob}), the two boundary points propagate through the kernel
$K_{\Delta}(t,z;x)$, with the intermediate bulk position integrated over.
The expression (\ref{massive celestial prop-final}) exhibits the same structure. The two celestial endpoints are connected through
the kernels $\mathcal{K}^{m}_{\Delta,h_i}(t;y,\mathbb{w};\mathbb{z}_i)$,
which are integrated over the intermediate point $(y,\mathbb{w})$ and over the
parameter $t$.

This observation allows us to identify the massive celestial
bulk-to-boundary kernel directly with the corresponding AdS
bulk-to-boundary kernel,
\begin{equation}
K_{\Delta}(t,z;x_i)
\quad\longrightarrow\quad
\mathcal{K}^{m}_{\Delta,h_i}(t;y,\mathbb{w};\mathbb{z}_i). \label{massive translation}
\end{equation}
An important feature of this correspondence is that the conformal dimension $\Delta$ and the AdS parameter $t$ appearing in
$\mathcal{K}^{m}_{\Delta}$ are not newly introduced celestial parameters. They are remnants of the original AdS propagator and survive the transformation to the celestial basis. In this translation, we build up the structural dictionary between AdS bulk-to-boundary propagator and celestial bulk-to-boundary kernel.


\section{Conclusions}
In this work, we studied how scalar propagators in AdS space are represented after transforming to the celestial basis. Beginning with the Euclidean AdS bulk-to-boundary propagator written in a Schwinger representation, we constructed the corresponding boundary-to-boundary propagator and subsequently mapped it to celestial space using conformal primary wavefunctions.

For the massless case, the celestial transformation leads to a
two-dimensional propagator on the celestial sphere. The direct
transformation initially produces a distributional two-point function
localized at coincident celestial points which is the Dirac delta function. After applying the shadow transformation, this distributional expression can be written in a conformally covariant form. We also express the celestial propagator analogous to the form of AdS/CFT boundary-to-boundary propagator. This allows us to structurally translate an AdS bulk-to-boundary propagator into a scaling rectified boundary-to-boundary celestial propagator (\ref{bulk boundary celestial translation}).

For the massive case, the celestial propagator exhibits a richer structure. In particular, the propagator naturally contains the modified Bessel function of the second kind, $K_{\Delta-2}(2m\sqrt{t})$, which resembles the radial dependence of momentum-space AdS bulk-to-boundary propagators. The comparison between the AdS and celestial representations reveals a
structural dictionary between their bulk-to-boundary propagators. An AdS bulk-to-boundary kernel is structurally translated to a celestial bulk-to-boundary propagator (\ref{massive translation}).

The results reveal a structural similarity between AdS gluing representations and celestial propagators which construct direct translations from AdS propagators, particularly bulk-to-boundary propagators, to celestial propagators.

Finally, the massive result smoothly connects to the massless one. In the limit $m\rightarrow0$, the Bessel-function factor reduces to one from the massless case. This provides a consistency check on the
construction and shows that the massive and massless celestial propagators can be understood within the same structural framework.

\section*{Acknowledgement}
We are grateful to the National Science, Research and Innovation Fund (NSRF) via the Program Management Unit for Human Resources \& Institutional Development, Research and Innovation for support under grant number B39G680009.

\appendix

\section{Decomposition of Dirac delta function} \label{dirac del}
We begin by decomposing the four-dimensional Dirac delta function $\delta^{(4)}(\mathbf{k} - \mathbf{k}')$ for massless particles. We start from the normalization condition
\begin{equation}
    \int d^4 \mathbf{k} \, \delta^{(4)}(\mathbf{k} - \mathbf{k}') = 1.
\end{equation}
We assume that the delta function admits the factorized form:
\begin{equation}
    \delta^{(4)}(\mathbf{k} - \mathbf{k}') = G(\mathbf{k} - \mathbf{k}') \, \delta(k^2)\, \Theta(k^0),
\end{equation}
where $\delta(k^2)\Theta(k^0)$ restricts the integration to the future light cone.

Using the standard identity,
\begin{equation}
    \int d^4 \mathbf{k} \, \delta(k^2)\Theta(k^0)\, f(\mathbf{k})
    = \int \frac{d^3 \vec{k}}{2|\vec{k}|} \, f(k^0 = |\vec{k}|, \vec{k}),
\end{equation}
we obtain
\begin{align}
    1 
    &= \int d^4 \mathbf{k} \, \delta(k^2)\Theta(k^0)\, G(\mathbf{k} - \mathbf{k}') \nonumber \\
    &= \int \frac{d^3 \vec{k}}{2|\vec{k}|} \, G(\mathbf{k} - \mathbf{k}').
    \label{eq:delta_identity}
\end{align}

Next, we parametrize the spatial momentum $\vec{k}$ in terms of $(\omega, z, \bar{z})$ as
\begin{equation}
    \vec{k}
    = \omega\bigl(z+\bar{z},\, -i(z-\bar{z}),\, 1 - |z|^2\bigr),
\end{equation}
which implies
\begin{equation}
    |\vec{k}| = \omega(1 + |z|^2).
\end{equation}
The corresponding Jacobian for this change of variables is
\begin{align}
    d^3 \vec{k}
    &= \left| \det \frac{\partial(k^1,k^2,k^3)}{\partial(\omega,z,\bar{z})} \right|
       d\omega\, dz\, d\bar{z} \nonumber \\
    &= 2\omega^2(1 + |z|^2)\, d\omega\, dz\, d\bar{z}.
\end{align}

Substituting into Eq.~\eqref{eq:delta_identity}, we find that the function $G(\mathbf{k}-\mathbf{k}')$ must take the form
\begin{equation}
    G(\mathbf{k} - \mathbf{k}') 
    = \frac{1}{\omega}\,
    \delta(\omega - \omega')\,
    \delta^{(2)}(z - z').
\end{equation}
Thus, the four-dimensional delta function for the massless case decomposes as
\begin{equation}
    \delta^{(4)}(\mathbf{k} - \mathbf{k}')
    = \frac{1}{\omega}\,
    \delta(\omega - \omega')\,
    \delta^{(2)}(z - z')\,
    \delta(k^2)\, \Theta(k^0).
\end{equation}

For a particle of mass $m$, we parametrize the four-momentum as
\begin{equation}
\mathbf{k}= m\,\mathbf{p}(y,z,\bar z),
\end{equation}
where $\mathbf{p}$ satisfies $p^2 = 1$ and is given by
\begin{equation}
\mathbf{p} =
\frac{1}{2y}
\left(
1+|z|^2+y^2,\;
z+\bar z,\;
-i(z-\bar z),\;
1-|z|^2-y^2
\right).
\end{equation}
The magnitude of the spatial momentum is then
\begin{equation}
    |\vec{p}|=\frac{1}{2y}\sqrt{(1+y^2+|z|^2)^2-4y^2}.
\end{equation}

The Lorentz-invariant phase space measure can be written as
\begin{equation}
d^4\mathbf{k}\,\delta(k^2-m^2)\theta(k^0)
=
m^2 \frac{d^3\vec{p}}{2\sqrt{|\vec{p}|^2+1}}.
\end{equation}

By calculating the Jacobian for the transformation to $(y,z,\bar{z})$ coordinates, we obtain
\begin{align}
    d^3\vec{p}&=\left| \det \frac{\partial(p^1,p^2,p^3)}{\partial(y,z,\bar{z})} \right|
       dy\, dz\, d\bar{z} \nonumber \\
    &= \frac{(1+y^2+|z|^2)}{4y^4}\, dy\, dz\, d\bar{z}.
\end{align}
Consequently, the measure simplifies to
\begin{equation}
\frac{d^3\vec{\hat{p}}}{2\sqrt{|\vec{p}|^2+1}}
=
\frac{1}{4 y^3} \, dy\, dz\, d\bar z.
\end{equation}

Following the same logic as the massless case, we define 
\begin{equation}
    \delta^{(4)}(\mathbf{k} - \mathbf{k}') = G(\mathbf{k} - \mathbf{k}') \, \delta(k^2-m^2)\, \Theta(k^0).
\end{equation}
To satisfy the normalization condition, $G(k-k')$ must satisfy
\begin{equation}
    G(\mathbf{k} - \mathbf{k}') 
    = \frac{4y^3}{m^2} \,
    \delta(y - y')\,
    \delta^{(2)}(z - z').
\end{equation}
We conclude that the massive four-dimensional delta function decomposes as
\begin{equation}
    \delta^{(4)}(\mathbf{k}-\mathbf{k}')=\frac{4y^3}{m^4} \delta(y - y')\,
    \delta^{(2)}(z - z') \delta(p^2-1)\Theta(p^0).
\end{equation}



\bibliographystyle{elsarticle-num} 
\bibliography{ref.bib}

@article{Pasterski:2017ylz,
    author = "Pasterski, Sabrina and Shao, Shu-Heng and Strominger, Andrew",
    title = "{Gluon Amplitudes as 2d Conformal Correlators}",
    eprint = "1706.03917",
    archivePrefix = "arXiv",
    primaryClass = "hep-th",
    doi = "10.1103/PhysRevD.96.085006",
    journal = "Phys. Rev. D",
    volume = "96",
    number = "8",
    pages = "085006",
    year = "2017"
}

@article{Pasterski:2017kqt,
    author = "Pasterski, Sabrina and Shao, Shu-Heng",
    title = "{Conformal basis for flat space amplitudes}",
    eprint = "1705.01027",
    archivePrefix = "arXiv",
    primaryClass = "hep-th",
    doi = "10.1103/PhysRevD.96.065022",
    journal = "Phys. Rev. D",
    volume = "96",
    number = "6",
    pages = "065022",
    year = "2017"
}

@article{Pasterski:2016qvg,
    author = "Pasterski, Sabrina and Shao, Shu-Heng and Strominger, Andrew",
    title = "{Flat Space Amplitudes and Conformal Symmetry of the Celestial Sphere}",
    eprint = "1701.00049",
    archivePrefix = "arXiv",
    primaryClass = "hep-th",
    doi = "10.1103/PhysRevD.96.065026",
    journal = "Phys. Rev. D",
    volume = "96",
    number = "6",
    pages = "065026",
    year = "2017"
}

@book{Gradshteyn:1943cpj,
    author = "Gradshteyn, I. S. and Ryzhik, I. M.",
    title = "{Table of Integrals, Series, and Products}",
    isbn = "978-0-12-294757-5, 978-0-12-294757-5",
    year = "1943"
}

@article{Ferrara:1972uq,
    author = "Ferrara, S. and Grillo, A. F. and Parisi, G. and Gatto, R.",
    title = "{The shadow operator formalism for conformal algebra. Vacuum expectation values and operator products}",
    doi = "10.1007/BF02907130",
    journal = "Lett. Nuovo Cim.",
    volume = "4S2",
    pages = "115--120",
    year = "1972"
}

@article{Maldacena:1997re,
    author = "Maldacena, Juan Martin",
    title = "{The Large $N$ limit of superconformal field theories and supergravity}",
    eprint = "hep-th/9711200",
    archivePrefix = "arXiv",
    reportNumber = "HUTP-97-A097, HUTP-98-A097",
    doi = "10.4310/ATMP.1998.v2.n2.a1",
    journal = "Adv. Theor. Math. Phys.",
    volume = "2",
    pages = "231--252",
    year = "1998"
}

@article{Witten:1998qj,
    author = "Witten, Edward",
    title = "{Anti de Sitter space and holography}",
    eprint = "hep-th/9802150",
    archivePrefix = "arXiv",
    reportNumber = "IASSNS-HEP-98-15",
    doi = "10.4310/ATMP.1998.v2.n2.a2",
    journal = "Adv. Theor. Math. Phys.",
    volume = "2",
    pages = "253--291",
    year = "1998"
}

@article{Aharony:1999ti,
    author = "Aharony, Ofer and Gubser, Steven S. and Maldacena, Juan Martin and Ooguri, Hirosi and Oz, Yaron",
    title = "{Large N field theories, string theory and gravity}",
    eprint = "hep-th/9905111",
    archivePrefix = "arXiv",
    reportNumber = "CERN-TH-99-122, HUTP-99-A027, LBNL-43113, RU-99-18, UCB-PTH-99-16, LBL-43113",
    doi = "10.1016/S0370-1573(99)00083-6",
    journal = "Phys. Rept.",
    volume = "323",
    pages = "183--386",
    year = "2000"
}

@article{Witten:1998zw,
    author = "Witten, Edward",
    editor = "Bergstrom, L. and Lindstrom, U.",
    title = "{Anti-de Sitter space, thermal phase transition, and confinement in gauge theories}",
    eprint = "hep-th/9803131",
    archivePrefix = "arXiv",
    reportNumber = "IASSNS-HEP-98-21",
    doi = "10.4310/ATMP.1998.v2.n3.a3",
    journal = "Adv. Theor. Math. Phys.",
    volume = "2",
    pages = "505--532",
    year = "1998"
}

@article{Ryu:2006bv,
    author = "Ryu, Shinsei and Takayanagi, Tadashi",
    title = "{Holographic derivation of entanglement entropy from AdS/CFT}",
    eprint = "hep-th/0603001",
    archivePrefix = "arXiv",
    reportNumber = "NSF-KITP-06-11, NSF-KITP-06-11",
    doi = "10.1103/PhysRevLett.96.181602",
    journal = "Phys. Rev. Lett.",
    volume = "96",
    pages = "181602",
    year = "2006"
}

@article{Strominger:1996sh,
    author = "Strominger, Andrew and Vafa, Cumrun",
    title = "{Microscopic origin of the Bekenstein-Hawking entropy}",
    eprint = "hep-th/9601029",
    archivePrefix = "arXiv",
    reportNumber = "HUTP-96-A002, RU-96-01",
    doi = "10.1016/0370-2693(96)00345-0",
    journal = "Phys. Lett. B",
    volume = "379",
    pages = "99--104",
    year = "1996"
}

@article{Hubeny:2007xt,
    author = "Hubeny, Veronika E. and Rangamani, Mukund and Takayanagi, Tadashi",
    title = "{A Covariant holographic entanglement entropy proposal}",
    eprint = "0705.0016",
    archivePrefix = "arXiv",
    primaryClass = "hep-th",
    reportNumber = "DCPT-07-13, KUNS-2069",
    doi = "10.1088/1126-6708/2007/07/062",
    journal = "JHEP",
    volume = "07",
    pages = "062",
    year = "2007"
}

@article{Gubser:1998bc,
    author = "Gubser, S. S. and Klebanov, Igor R. and Polyakov, Alexander M.",
    title = "{Gauge theory correlators from noncritical string theory}",
    eprint = "hep-th/9802109",
    archivePrefix = "arXiv",
    reportNumber = "PUPT-1767",
    doi = "10.1016/S0370-2693(98)00377-3",
    journal = "Phys. Lett. B",
    volume = "428",
    pages = "105--114",
    year = "1998"
}

@article{Policastro:2001yc,
    author = "Policastro, G. and Son, Dan T. and Starinets, Andrei O.",
    title = "{The Shear viscosity of strongly coupled N=4 supersymmetric Yang-Mills plasma}",
    eprint = "hep-th/0104066",
    archivePrefix = "arXiv",
    reportNumber = "NYU-TH-01-04-02, SNS-PH-01-05",
    doi = "10.1103/PhysRevLett.87.081601",
    journal = "Phys. Rev. Lett.",
    volume = "87",
    pages = "081601",
    year = "2001"
}

@article{Hartnoll:2009sz,
    author = "Hartnoll, Sean A.",
    editor = "Uranga, A. M.",
    title = "{Lectures on holographic methods for condensed matter physics}",
    eprint = "0903.3246",
    archivePrefix = "arXiv",
    primaryClass = "hep-th",
    doi = "10.1088/0264-9381/26/22/224002",
    journal = "Class. Quant. Grav.",
    volume = "26",
    pages = "224002",
    year = "2009"
}

@article{Polyakov:1998ju,
    author = "Polyakov, Alexander M.",
    title = "{The Wall of the cave}",
    eprint = "hep-th/9809057",
    archivePrefix = "arXiv",
    reportNumber = "PUPT-1812",
    doi = "10.1142/S0217751X99000324",
    journal = "Int. J. Mod. Phys. A",
    volume = "14",
    pages = "645--658",
    year = "1999"
}

@article{Srisangyingcharoen:2025ett,
    author = "Srisangyingcharoen, Pongwit and Pukdee, Jongruk",
    title = "{Two-point tree-level string amplitudes as AdS transition amplitudes}",
    eprint = "2505.07358",
    archivePrefix = "arXiv",
    primaryClass = "hep-th",
    doi = "10.1016/j.physletb.2025.139736",
    journal = "Phys. Lett. B",
    volume = "868",
    pages = "139736",
    year = "2025"
}

@article{Alday:2007hr,
    author = "Alday, Luis F. and Maldacena, Juan Martin",
    title = "{Gluon scattering amplitudes at strong coupling}",
    eprint = "0705.0303",
    archivePrefix = "arXiv",
    primaryClass = "hep-th",
    reportNumber = "SPIN-07-16, ITP-UU-07-24",
    doi = "10.1088/1126-6708/2007/06/064",
    journal = "JHEP",
    volume = "06",
    pages = "064",
    year = "2007"
}

@article{Penedones:2010ue,
    author = "Penedones, Joao",
    title = "{Writing CFT correlation functions as AdS scattering amplitudes}",
    eprint = "1011.1485",
    archivePrefix = "arXiv",
    primaryClass = "hep-th",
    doi = "10.1007/JHEP03(2011)025",
    journal = "JHEP",
    volume = "03",
    pages = "025",
    year = "2011"
}

@article{Fitzpatrick:2011ia,
    author = "Fitzpatrick, A. Liam and Kaplan, Jared and Penedones, Joao and Raju, Suvrat and van Rees, Balt C.",
    title = "{A Natural Language for AdS/CFT Correlators}",
    eprint = "1107.1499",
    archivePrefix = "arXiv",
    primaryClass = "hep-th",
    reportNumber = "SLAC-PUB-14506, HRI-ST-1107",
    doi = "10.1007/JHEP11(2011)095",
    journal = "JHEP",
    volume = "11",
    pages = "095",
    year = "2011"
}

@article{Liu:1998ty,
    author = "Liu, Hong and Tseytlin, Arkady A.",
    title = "{On four point functions in the CFT / AdS correspondence}",
    eprint = "hep-th/9807097",
    archivePrefix = "arXiv",
    reportNumber = "IMPERIAL-TP-97-98-060",
    doi = "10.1103/PhysRevD.59.086002",
    journal = "Phys. Rev. D",
    volume = "59",
    pages = "086002",
    year = "1999"
}

@article{Costa:2014kfa,
    author = "Costa, Miguel S. and Gon{\c{c}}alves, Vasco and Penedones, Jo{\~a}o",
    title = "{Spinning AdS Propagators}",
    eprint = "1404.5625",
    archivePrefix = "arXiv",
    primaryClass = "hep-th",
    doi = "10.1007/JHEP09(2014)064",
    journal = "JHEP",
    volume = "09",
    pages = "064",
    year = "2014"
}

@article{Gopakumar:2003ns,
    author = "Gopakumar, Rajesh",
    title = "{From free fields to AdS}",
    eprint = "hep-th/0308184",
    archivePrefix = "arXiv",
    doi = "10.1103/PhysRevD.70.025009",
    journal = "Phys. Rev. D",
    volume = "70",
    pages = "025009",
    year = "2004"
}

@article{Gopakumar:2004qb,
    author = "Gopakumar, Rajesh",
    title = "{From free fields to AdS. 2.}",
    eprint = "hep-th/0402063",
    archivePrefix = "arXiv",
    doi = "10.1103/PhysRevD.70.025010",
    journal = "Phys. Rev. D",
    volume = "70",
    pages = "025010",
    year = "2004"
}

@article{Weinberg:1965nx,
    author = "Weinberg, Steven",
    title = "{Infrared photons and gravitons}",
    doi = "10.1103/PhysRev.140.B516",
    journal = "Phys. Rev.",
    volume = "140",
    pages = "B516--B524",
    year = "1965"
}

@article{Strominger:2013jfa,
    author = "Strominger, Andrew",
    title = "{On BMS Invariance of Gravitational Scattering}",
    eprint = "1312.2229",
    archivePrefix = "arXiv",
    primaryClass = "hep-th",
    doi = "10.1007/JHEP07(2014)152",
    journal = "JHEP",
    volume = "07",
    pages = "152",
    year = "2014"
}

@article{He:2014laa,
    author = "He, Temple and Lysov, Vyacheslav and Mitra, Prahar and Strominger, Andrew",
    title = "{BMS supertranslations and Weinberg{\textquoteright}s soft graviton theorem}",
    eprint = "1401.7026",
    archivePrefix = "arXiv",
    primaryClass = "hep-th",
    doi = "10.1007/JHEP05(2015)151",
    journal = "JHEP",
    volume = "05",
    pages = "151",
    year = "2015"
}

@article{Kapec:2014opa,
    author = "Kapec, Daniel and Lysov, Vyacheslav and Pasterski, Sabrina and Strominger, Andrew",
    title = "{Semiclassical Virasoro symmetry of the quantum gravity $ \mathcal{S}$-matrix}",
    eprint = "1406.3312",
    archivePrefix = "arXiv",
    primaryClass = "hep-th",
    doi = "10.1007/JHEP08(2014)058",
    journal = "JHEP",
    volume = "08",
    pages = "058",
    year = "2014"
}

@article{Arkani-Hamed:2020gyp,
    author = "Arkani-Hamed, Nima and Pate, Monica and Raclariu, Ana-Maria and Strominger, Andrew",
    title = "{Celestial amplitudes from UV to IR}",
    eprint = "2012.04208",
    archivePrefix = "arXiv",
    primaryClass = "hep-th",
    doi = "10.1007/JHEP08(2021)062",
    journal = "JHEP",
    volume = "08",
    pages = "062",
    year = "2021"
}

@article{Stieberger:2018edy,
    author = "Stieberger, Stephan and Taylor, Tomasz R.",
    title = "{Strings on Celestial Sphere}",
    eprint = "1806.05688",
    archivePrefix = "arXiv",
    primaryClass = "hep-th",
    reportNumber = "MPP-2018-136",
    doi = "10.1016/j.nuclphysb.2018.08.019",
    journal = "Nucl. Phys. B",
    volume = "935",
    pages = "388--411",
    year = "2018"
}

@article{Taylor:2023bzj,
    author = "Taylor, Tomasz R. and Zhu, Bin",
    title = "{Celestial Supersymmetry}",
    eprint = "2302.12830",
    archivePrefix = "arXiv",
    primaryClass = "hep-th",
    doi = "10.1007/JHEP06(2023)210",
    journal = "JHEP",
    volume = "06",
    pages = "210",
    year = "2023"
}

@article{Donnay:2020guq,
    author = "Donnay, Laura and Pasterski, Sabrina and Puhm, Andrea",
    title = "{Asymptotic Symmetries and Celestial CFT}",
    eprint = "2005.08990",
    archivePrefix = "arXiv",
    primaryClass = "hep-th",
    reportNumber = "CPHT-RR022.042020",
    doi = "10.1007/JHEP09(2020)176",
    journal = "JHEP",
    volume = "09",
    pages = "176",
    year = "2020"
}

@article{Puhm:2019zbl,
    author = "Puhm, Andrea",
    title = "{Conformally Soft Theorem in Gravity}",
    eprint = "1905.09799",
    archivePrefix = "arXiv",
    primaryClass = "hep-th",
    reportNumber = "CPHT-RR021.052019",
    doi = "10.1007/JHEP09(2020)130",
    journal = "JHEP",
    volume = "09",
    pages = "130",
    year = "2020"
}

@article{Yuenyong:2026wnv,
    author = "Yuenyong, Aphiwat and Srisangyingcharoen, Pongwit and Hirunsirisawat, Ekapong and Deesuwan, Tanapat",
    title = "{Integral Transformations for Conformally Invariant Celestial Amplitudes}",
    eprint = "2602.14422",
    archivePrefix = "arXiv",
    primaryClass = "hep-th",
    month = "2",
    year = "2026"
}

@article{Furugori:2023hgv,
    author = "Furugori, Hideo and Ogawa, Naoki and Sugishita, Sotaro and Waki, Takahiro",
    title = "{Celestial two-point functions and rectified dictionary}",
    eprint = "2312.07057",
    archivePrefix = "arXiv",
    primaryClass = "hep-th",
    reportNumber = "KUNS-2988, YITP-23-160",
    doi = "10.1007/JHEP02(2024)063",
    journal = "JHEP",
    volume = "02",
    pages = "063",
    year = "2024"
}


\end{document}